\documentclass[aps,preprint,superscriptaddress,groupedaddress]{revtex4}  
\usepackage{graphicx}
\usepackage{dcolumn}   
\usepackage{bm}        
\usepackage{amssymb}   
\usepackage{amsmath}
\usepackage[x11names]{xcolor}

\newcommand{\colr}{\color{black}}

\newcommand{\dif}{\partial}
\newcommand{\zf}{V_{\rm ZF}}
\newcommand{\f}{Z}
\newcommand{\qx}{{q_r}}

\newcommand{\ky}{k_\theta}
\newcommand{\gk}{\gamma_k}

\newcommand{\stair}{\Delta}

\newcommand{\upg}{\nu_{\rm UG}}
\newcommand{\hyp}{\nu_h}
\newcommand{\nup}{{\nu_\parallel}}

\begin{document}

\title{Friction-induced scale-selection in the extended Cahn-Hilliard model for zonal staircase}

\author{M. Leconte}
\affiliation{Korea Institute of Fusion Energy (KFE), Daejeon 34133, South Korea}
\author{T.S. Hahm}
\affiliation{Nuclear Research Institute for Future Technology and Policy, Seoul National University, Seoul 08826, South Korea}

\begin{abstract}
{\colr In this work, we describe a possible mechanism to set the radial scale of zonal flows, which may be applicable to the $E \times B$ staircase found in the global full-f simulations such as [G. Dif-Pradalier et al. \emph{Phys. Rev. Lett.} 114, 085004 (2015)]}. 1D numerical simulation results of the Cahn-Hilliard model - extended to include zonal flow friction - can be understood from a heuristic nonlinear analysis. The staircase step-size $\Delta$ is found to decrease as the dimensionless zonal flow friction $\mu$ increases. It scales like $\Delta \sim \log \mu^\alpha$, with $\alpha \simeq -0.41$, up to a constant offset.
\end{abstract}

\maketitle

Pattern formation occurs in many far-from-equilibrium physical and biological systems. It involves self-organized nonlinear structures \cite{CrossGreenside2009}. Radially sheared $E \times B$ zonal flows (ZF) are the prime example of such in magnetized plasmas, and can regulate turbulence and transport \cite{DiamondI2Hahm2005}. They form coherent radial bands known as 'staircase'. Therefore, understanding this staircase structure is an outstanding scientific issue with a high fusion relevance. In particular, the $E \times B$ staircase has been observed from both gyrokinetic simulations \cite{DifPradalier2015,DifPradalier2017,WWang2018,LeiQi2022} and tokamak experiments \cite{Hornung2017,MJChoi2024}. While there has been steady theoretical progress \cite{KosugaDiamondGurcan2013,AshourvanDiamond2017,Sasaki2021,LeconteKobayashi2021,GarbetPanicoVarennes2021, Gruzinov1994} in understanding the formation mechanism, its dependence on key physical parameters has not been investigated in detail to date. In particular, the influence of frictional damping of zonal flows on the $E \times B$ staircase spacing, or \emph{step-size} has not been addressed. This information is useful in extrapolating predictions to future machines including a possibility of internal barrier (ITB) formation. Moreover, it can serve as a critical test for the validity of models. \newline
\indent While there exists a variety of theoretical models for the staircase, one of the simplest - but highly relevant to many physical systems including phase separation, i.e. spinodal decomposition in binary solids and fluids also called Ostwald ripening \cite{OonoBahiana1988,Villain-Guillot2010}, geophysical fluid dynamics (GFD) \cite{ManfroiYoung1999}, and magnetic fusion energy (MFE) \cite{Itoh2004,Itohppcf2004,Itoh2005} - is the Cahn-Hilliard equation (CH) or numerous extensions thereof. In the magnetic fusion energy (MFE) context, it can be derived by a formal expansion of the wave-kinetic equation in terms of the zonal flow shear magnitude \cite{Itoh2004}.
In this Letter, we show that the extended Cahn-Hilliard model including zonal flow friction - derived from the wave kinetic equation - can predict the step-size - i.e. radial scale - of the staircase. The \emph{novelty} lies in the synergy between the \emph{nonlinearity} and \emph{dissipation} due to ZF friction, which allows for a finite ZF scale. \newline
\indent Using a heuristic nonlinear analysis, we estimate the $\mu$-dependence of the staircase step size where $\mu$ is the ZF friction defined after Eq.(\ref{ch00}).
This yields the following scaling law for the staircase step-size, Eq.(\ref{scaling-ana}): $\stair_{\rm stair} \sim 
\log \mu^{\beta} + {\rm const}$, with $\beta \simeq -0.34$. {\colr Here, and throughout the text, $\log=\log_{10}$ denotes the base-10 logarithm}. This analytically-derived scaling law is then compared to a 1D numerical simulation of the model which yields the following scaling law, Eq.(\ref{scaling-num}): $\stair_{\rm stair} \sim
\log \mu^\alpha + {\rm const}$ , with scaling exponent $\alpha \simeq -0.41$, showing a good agreement. Eqs.(\ref{scaling-ana}) and (\ref{scaling-num}) are the main results of this Letter. \newline
\indent \emph{Model.$-$}The wave-kinetic equation \cite{DiamondI2Hahm2005} describes the distortion of a drift-wave wave-packet having wave-action density $N_k = (1+k_r^2+\ky^2)^2|\phi_k|^2$, due to $E \times B$ zonal flows $V_{\rm ZF} = \dif_r \langle \phi \rangle$ with $\phi$ the electric potential and $\langle \cdot \rangle$ a zonal average. Here, $(r,\theta)$ are the radial and poloidal directions in a tokamak, and ($k_r,\ky$) are the associated wavenumbers in units of $\rho_s$, the ion gyroradius at electron temperature. We will also use the local slab variable $x=r-r_{\rm ref}$, with $r_{\rm ref}$ a reference radius. \newline
\indent The wave-kinetic equation (WKE) takes the form: $\dif_t N_k + (\dif_{k_r} \omega_k) \dif_r N_k - \dif_r (\ky \zf) \dif_{k_r} N_k = 2 \gk N_k + C$, where $\gk$ denotes the linear growth-rate - the form of which depends on the turbulence model - and the 'collision integral' $C$ is taken as $C= - \Delta \omega N_k^2$. Here, $\omega_k$ is the drift-wave frequency and $\Delta \omega$ denotes the turbulent decorrelation-rate.
The quantities $N_k$ and $\zf$ are written in Fourier series: $(N_k,\zf) = \sum_q (N_q,V_q) e^{i \qx x - i \Omega t} +c.c.$, with $\qx$ the zonal wavenumber and $\Omega$ the zonal frequency. Assuming the balance $2 \gk N_k - \Delta\omega N_k^2 \simeq 0$ is approximately satisfied, the WKE yields at lowest order a mixing-length estimate for the background wave action density: $N_{k0} = 2 \gk / \Delta\omega $. After some algebra, the first-order modulation of turbulence by zonal flows reads $N_q^{(1)} = \ky R_q V'_q \frac{\dif \langle N_k \rangle }{\dif k_r}$, where $V'_q = i \qx V_q$, $R_q = 1 / (-i(\Omega - \qx v_g) + 2\gk) \sim 1 / 2 \gk$ for broadband turbulence, and $v_g = d \omega_k / dk_r$ is the radial group velocity. The second-order modulation does not contribute to zonal flow evolution when summing over $k_\theta$, due to the system periodicity in  $\theta$. The third-order modulation is then given by: $N_q^{(3)} = - \ky^3 R_{2q} |R_q|^2 |V'_q|^2 V'_q \frac{\dif^3 \langle N_k \rangle}{\dif k_r^3}$ \cite{Itoh2004,Itoh2005,Itohppcf2004}, where $|\cdot|^2$ denotes the square modulus. \newline
\indent The zonal flow evolution is then given by $\dif_t \zf = - \dif_x [\Pi + \Pi_\parallel] - \hat \mu \zf$, with $\Pi = \langle \tilde v_r \tilde v_ \theta \rangle \sim \sum_{k,\qx} k_r \ky [N_q e^{i \qx x} + c.c.] / (1+k_r^2+ \ky^2)^2$ the Reynolds stress, $\Pi_\parallel = - \nup \dif_x \zf$ the parallel stress and $\hat \mu$ the zonal flow neoclassical friction. The latter is given by: $\hat \mu = 0.67 \nu_{ii} / \epsilon$, with $\nu_{ii}$ the ion collision frequency and $\epsilon=a/R$ the inverse aspect-ratio, with $a$ the minor radius and $R$ the major radius \cite{HintonRosenbluth1999}. Replacing the first and third order contributions into the zonal flow evolution, one obtains the following extended Cahn-Hilliard equation for the \emph{zonal flow shear} $\f = \dif_x \zf$, written in variational form:
$
\dif_t \f = \dif_{xx} \Big[ \frac{\delta F_{\rm GL}(\f(x,t))}{\delta \f(x,t)} \Big] - \hat \mu \f,
$
where $F_{\rm GL} = \int \upg \Big[ -\frac{\f^2}{2} + \frac{\f^4}{4} \Big] - \frac{1}{2} \hyp (\dif_x \f)^2 dx$ denotes the Ginzburg-Landau \emph{free energy} \cite{Gruzinov1994,Villain-Guillot2010}. This corresponds to a mean-field approach.
The equivalent non-dimensional form is:
\begin{equation}
\dif_t \f = - \dif_{xx} [ (1- \nup) \f - \f^3] - \dif_{xxxx} \f - \mu \f,
\label{ch00}
\end{equation}
where the radial variable is normalized as $\frac{x}{L} \to x$, with $L= \delta_h$. In Eq.(\ref{ch00}), $\mu = \mu \tau_h$ is the dimensionless ZF friction, proportional to the ratio of the ion-ion collision rate to the ZF modulational growth-rate. Typical estimates range from $\mu \sim 1.7\times 10^{-4}$ for ITER to $\mu \sim 3.5 \times 10^{-3}$ for KSTAR. Time is normalized as $t / \tau_h \to t$ with $ \tau_h = \frac{\delta_h^2}{\upg} $ proportional to the hyperviscous time $\delta_h^2 / \upg$. The parameter $\nu_\parallel = \nu_\parallel^{\rm phys} / \upg$ is the normalized - parallel - turbulent shear viscosity, which provides the ZF threshold in the friction-less limit $\mu=0$. Here, $\delta_h$ denotes the width of the hyper-viscous layer given by $\delta_h = \sqrt{\hyp / \upg}$, where $\upg = \frac{1}{2} \frac{\dif^2 \gamma_q}{\dif \qx^2} \Big|_{\qx^0}$ is the ZF up-gradient viscosity, related to the familiar effective 'negative' viscosity $\nu_{\rm eff} = - \upg \le 0$, with $\qx^0$ a reference ZF wavenumber and $\gamma_q \sim I_0 \alpha_q \qx^2$ the ZF modulational growth-rate, $I_0$ the reference turbulence intensity without ZFs, $\alpha_q$ the turbulence-ZF coupling parameter \cite{DiamondI2Hahm2005}. For a given $\nup$, the parameter $\mu = \hat \mu \tau_h$ is the only dimensionless parameter.
While Eq.(\ref{ch00}), on surface, may look like describing a self-interaction of zonal flows which is theoretically impossible for the $E\times B$ vector nonlinearity \cite{Diamond-IAEA1998}, it should be noted that turbulence intensity enters Eq.(\ref{ch00}) via the ZF up-gradient viscosity. So Eq.(\ref{ch00}) describes zonal flow - zonal flow interaction mediated by microturbulence. \newline
\indent We note that, based on mixing-length estimate, the wave-kinetic equation implies that $I_0$ - and hence $\upg$ - increases with the distance from marginal stability, namely $I_0 \propto \sum \gamma_k / \Delta\omega$.
Based on this estimate, one can see that near marginality, defined as $\sum \gamma_k \sim 0$, the up-gradient diffusivity becomes very small. For typical ion-temperature gradient-driven (ITG) turbulence, linear theory yields $\gamma_k \propto \sqrt{\frac{R}{L_T}-\frac{R}{L_T^c}}$, with $L_T = \Big| \frac{\nabla T}{T} \Big|^{-1}$ the temperature-gradient scale length and $L_T^c$ its linear critical value \cite{DifPradalierPRE2010}. Hence, the characteristic ZF scale $\sim \delta_h \propto 1 / \sqrt{\upg}$ is predicted to increase close to marginality $\frac{R}{L_T} \simeq \frac{R}{L_T^c}$. This is consistent with Ref.\cite{DifPradalier2017}. In Fig.4 of this Reference, it is shown that the PDF of staircase corrugations v.s. driving gradient ($\frac{R}{L_T}$) is localized  near marginality, although marginality there is defined with respect to the nonlinear critical gradient.

\begin{figure}
\includegraphics[width=0.5\linewidth]{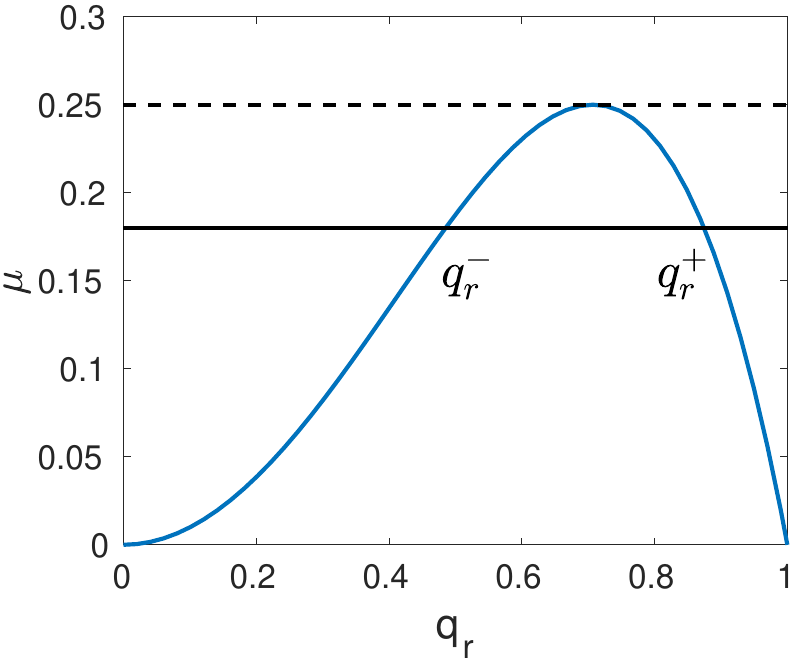}
\caption{The linear dispersion relation, Eq.(\ref{lindisp}) is shown. The black horizontal line shows that for a given value of the zonal flow friction $\mu$, there are two possible radial wavenumbers: $\qx_+$ and $\qx_-$. The dashed line shows the maximal value of friction $\mu_c = \frac{1}{4}$, above which zonal flows are totally suppressed.}
\label{fig:intro}
\end{figure}

\begin{figure}
\begin{center}
\includegraphics[width=0.5\linewidth]{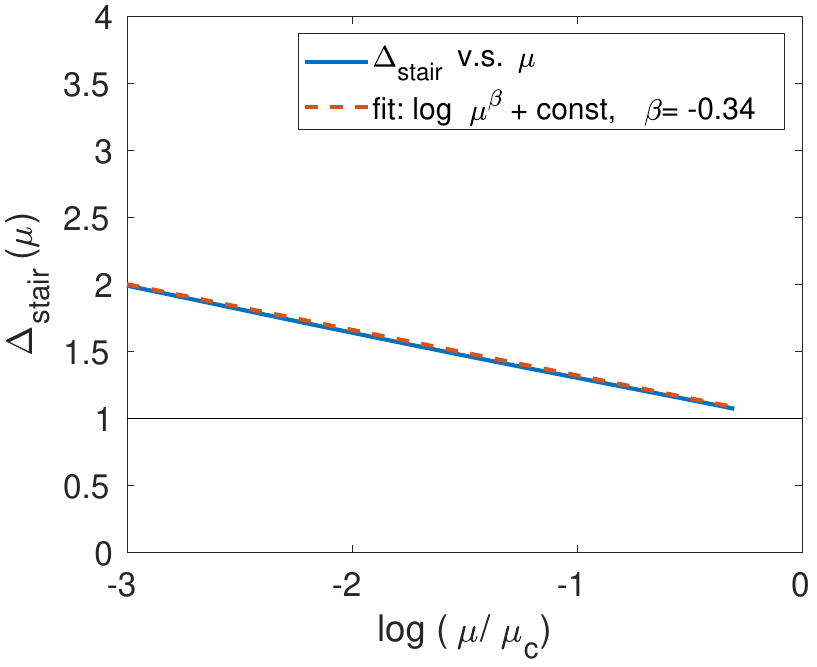}
\caption{The scale-selection is shown graphically: Semi-log plot of staircase step-size $\stair_{\rm stair}$ v.s. ZF friction $\mu$v.s. ZF friction $\mu$ (solid-blue)\colr{ evaluated using the formula of Ref.\cite{formula-sheet} at 2nd order in $1-\kappa(\mu)$}. The dashed-line is a semi-log fit.}
\label{fig:fig2}
\end{center}
\end{figure}

\begin{figure}
\includegraphics[width=0.5\linewidth]{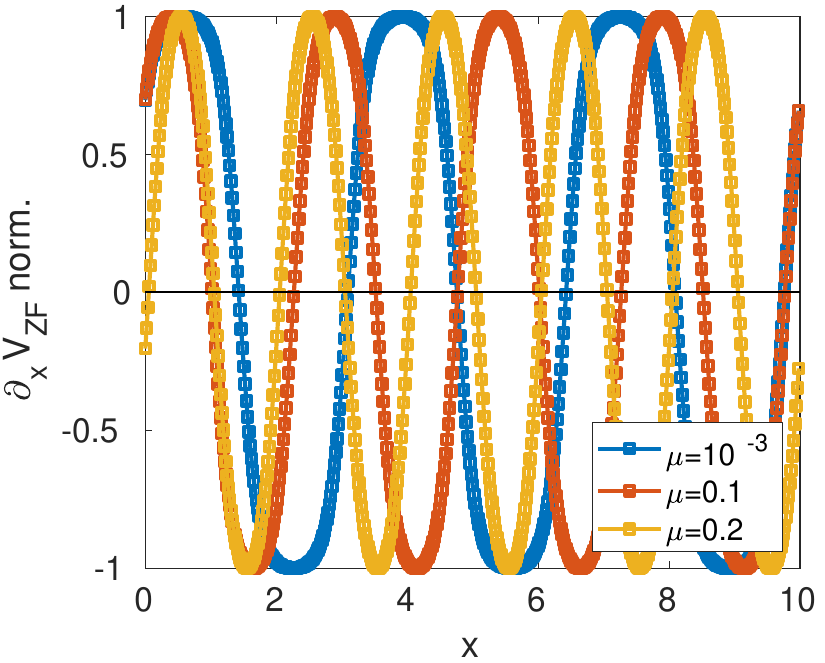}
\caption{The time-averaged zonal flow shear profile $Z=\dif_x \zf$ - normalized to its amplitude - is shown for different values of ZF friction $\mu=10^{-3}$ (blue), $\mu=0.1$ (red) and $\mu=0.2$ (yellow).}
\label{fig:fig3}
\end{figure}

\begin{figure}
\includegraphics[width=0.5\linewidth]{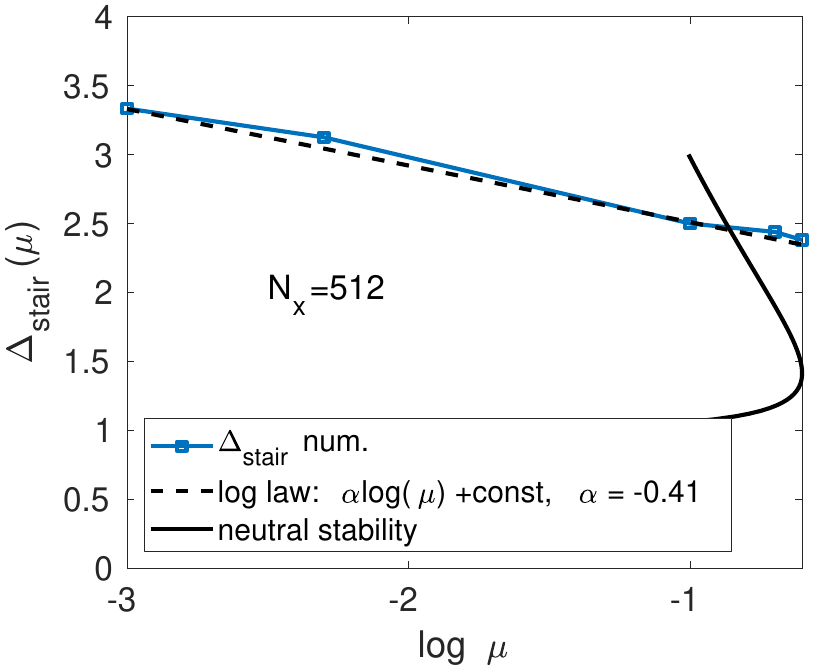}
\caption{Staircase nonlinear dispersion relation: Semi-log plot of $\Delta_{\rm stair}$ v.s. ZF friction $\mu$ from the 1D numerical simulation for $N_x=512$ radial points. The dashed-line is a semi-log fit of $\stair_{\rm stair}$. The black solid-line shows the neutral-stability curve: $\frac{1}{\qx_\pm}$ v.s. $\mu$.}
\label{fig:fig4}
\end{figure}

Note that Eq.(\ref{ch00}) is similar to the equation derived for atmospheric zonal jets in Ref.\cite{ManfroiYoung1999} in the geophysical fluid dynamics context (GFD), and to the zonal equation derived from a formal expansion of the wave-kinetic equation \cite{Itoh2004}.
Physically, in the present model, the hyper-viscosity term $\propto \hyp$ can be related to
residual short-scale turbulence. The associated 'smoothing effect' is strong with a high-order derivative, while the hyper-viscosity itself is small in magnitude. Here,  the zonal flow shear and up-gradient diffusivity $\upg$ are normalized to their maximal value such that: $ \sqrt{ \frac{1+ \kappa^2}{2 \kappa^2 } } \f \to \f$, and $(1+ \kappa^2) \upg \to \upg$ with $\kappa$ the Jacobi elliptic modulus. Physically, $\kappa$ is related to the turbulence parameters via $\frac{2 \kappa^2}{1+ \kappa^2} \sim H \frac{\ky^2}{ (\Delta \omega)^2 }$, with $H = H(\qx, \Delta\omega) \sim 2.5$ a coefficient \cite{Itoh2005}. Hence, the ZF shear magnitude essentially scales with the turbulent decorrelation rate $|Z_{\rm max}| \sim \Delta\omega$, \emph{valid in the regime of large turbulent decorrelation} $\Delta\omega$. In the opposite regime of small $\Delta\omega$, the resonance function is modified to $R_q \sim \frac{1}{\Gamma}$, with the collisionless damping $\Gamma \propto \omega_b$ proportional to the bounce frequency: $\omega_b \propto \sqrt{\qx |Z|}$. Hence, the 3rd-order truncation of the WKE, and associated Cahn-Hilliard equation is still valid in the regime of small $\Delta\omega$, provided the ordering $\ky^2 H Z^2 < \omega_b^2$ is satisfied. This requires $\Delta_{\rm stair} \propto \frac{1}{\qx}$ not too large. \newline
\indent In the reference case of the standard Cahn-Hilliard equation, $\mu \to 0$, at saturation $\dif_t=0$, Eq.(\ref{ch00}) admits the Jacobi elliptic sine '${\rm sn}$' as nonlinear solution \cite{Itoh2005}:
\begin{equation}
\f_{\rm sat}(x) \to \f_{\rm max} (\kappa) {\rm sn}(x, \kappa).
\end{equation}
 This friction-less solution has a sinusoidal shape at small amplitude ($\kappa \ll 1$), while it has a square-wave like shape at large amplitude ($\kappa \simeq 1$) as discussed in Ref.\cite{Itoh2005}. It should be noted that time-dependent simulations indicate a trend that the scale becomes as large as allowed by the system in time \cite{ManfroiYoung1999, Obuse2011}, with a logarithmic time dependence \cite{PolitiMisbah2006}. The effect of very small ZF friction has been investigated by treating that term as a pertubation to the frictionless solution \cite{Itoh2004,Itohppcf2004}. In this work, we seek a solution for which every term on the RHS of Eq.(\ref{ch00}) is taken on an equal footing. This should extrapolate better to a wider range of friction values.
We are primarily interested in the steady-state version of Eq.(\ref{ch00}), and we set $\nu_\parallel=0$ since we focus on the effect of ZF friction:
\begin{equation}
\dif_{xx} [ \f - \f^3 ] + \dif_{xxxx} \f + \mu \f = 0,
\label{ch0}
\end{equation}
\indent \emph{Linear analysis.$-$}Linearizing Eq.(\ref{ch0}) around $\f=0$ yields the following relation:
\begin{equation}
\qx^4 - \qx^2 + \mu = 0.
\label{lindisp}
\end{equation}
Defining the characteristic scale associated to the ZF friction $\delta_\mu = 1 /\sqrt{\mu}$, we note that $\delta_\mu \to + \infty$ in the limit of vanishing ZF friction, so $\delta_\mu$ acts as an infrared cut-off scale. The standard Cahn-Hilliard equation $(\mu=0)$ has no such cut-off, so in this case, the flow pattern reaches the largest scale possible - the domain size - in finite time. The solution to Eq.(\ref{lindisp}) is $\qx_\pm = \frac{1}{\sqrt{2}} \sqrt{ 1 \pm \sqrt{ 1 -4 \mu}}$, where only the positive branch $\qx \ge 0$ is displayed for clarity. The $\pm$ superscript denotes the two sub-branches, corresponding to the $\pm$ sign in the square-root. This relation is shown [Fig.\ref{fig:intro}]. Hence, there exists a \emph{critical} ZF friction $\mu_c$ above which the relation admits no radially-periodic solution, and thus $\f=0$ is the only solution of Eq.(\ref{ch00}). Therefore $\mu \le \mu_c$ is required for a staircase solution $\f \neq 0$. The critical ZF friction is $\mu_c = \frac{1}{4}$.
Hence, one can get some insight about the effect of ZF friction $\mu$ on the ZF radial scale, even in the linear regime: The ZF wavenumber $\qx$ for the relevant longer wavelength branch \emph{increases} - i.e. the ZF radial scale $2\pi / \qx$ \emph{decreases} - when the ZF friction $\mu$ increases. Also, ZF friction is constrained by the marginal stability condition $\mu \le \mu_c = \frac{1}{4}$. For higher values of ZF friction, zonal flows are completely suppressed. \newline
\indent \emph{Nonlinear Analysis.$-$ 
Since a rigorous analytical solution of Eq.(3) is beyond the scope of this paper even if not impossible,
we pursue the following heuristic approach. We make an ansatz in which the cubic nonlinear term $Z^3$ in Eq.(3)
is approximated by $Z_s^2 \cdot Z$ where $x$-independent $Z_s$ quantitifies the zonal shear amplitude.
Then, we obtain:
\begin{equation}
\qx^4 -(1-Z_s^2)\qx^2 + \mu = 0,
\end{equation}
which extends Eq.(\ref{lindisp}) to the nonlinear regime. Then, the relevant low-wavenumber branch solution of Eq.(5) satisfies
\begin{equation}
\qx_-^2 = \frac{1}{2} \Big[ (1-Z_s^2) - \sqrt{ (1- Z_s^2)^2 -4 \mu  } \Big]
\end{equation}
Eq.(6) indicates that now the critical value for $\mu$ is reduced to $\frac{1}{4}(1-Z_s^2)$
and $\qx_-$ increases as the zonal shear amplitude $Z_s$ increases in the nonlinear regime. A useful observation for
the staircase width estimation is that $Z_s^2$ value is bounded by $1-2 \sqrt{\mu}$ from above, since
$1-Z_s^2 = \qx^2 + \frac{\mu}{\qx^2} \ge 2 \sqrt{\mu}$.
For small $\mu$, by equating $Z_s^2$ to $Z_{\rm max}^2 = \frac{2 \kappa^2}{1 + \kappa^2}$, in the Jacobi-elliptic solution of the fully nonlinear equation in Eq.(2), we can relate $\kappa$ (the elliptic modulus) to $\mu$ (the dimension-less friction).
Then, we obtain $\kappa = 1 - \sqrt{ \frac{\mu}{\mu_c} }$ with $\mu_c = \frac{1}{4}$, and the desired expression for the staircase step-size is given as
\begin{equation}
\Delta_{\rm stair} \simeq \frac{2}{\pi} K(\kappa), \quad {\rm where} \quad \kappa = 1- \sqrt{ \frac{\mu}{\mu_c} }
\label{nldisp}
\end{equation}
For $\mu \ll \mu_c$, an asymptotic formula {\colr of $K(\kappa)$ for $\kappa \sim 1$, up to the 2nd order in $1-\kappa$ \cite{formula-sheet}, modified to the base-10 logarithm}, then directly yields:
\begin{equation}
\Delta_{\rm stair} \simeq \beta \log \Big[ \frac{\mu}{\mu_c} \Big] + {\rm const}, \quad {\rm with}
\quad \beta \simeq - 0.34
\label{scaling-ana}
\end{equation}
}
The elliptic modulus and associated staircase step-size (Eq.\ref{nldisp}) is shown v.s. friction $\mu$ [Fig.\ref{fig:fig2}].
This semi-log law arises because the step-size depends on the elliptic integral $K(\kappa)$. This integral diverges as $\kappa$ approaches $1$, and hence does not follow a power-law.
This scaling is different from the results of Refs.\cite{OonoBahiana1988, LiuGoldenfeld1989} obtained in the context of spinodal decomposition in metal alloys and block-copolymer melts, respectively, which obtained an algebraic dependence of the step-size on the ZF friction.
For example, Ref.\cite{LiuGoldenfeld1989} reports scaling laws $\Delta \sim \mu^{- \gamma}$, with $\gamma \simeq 0.3$ which is similar to Ref.\cite{ManfroiYoung1999} who found $\gamma \simeq \frac{1}{3}$. \newline
\indent \emph{Numerical results.$-$} Since we are mostly interested in $\qx L <1$ for ZFs, where $L \sim \delta_h$, the following ordering is appropriate: $1 \le \delta_h \le \delta_\mu$. Eq.(\ref{ch00}) is solved using a 1D spectral code with semi-implicit scheme, including anti-aliasing, in a spatial domain with $N_x=512$ grid points, and size $x = [0,10]$ with a periodic boundary condition. The initial condition used is a sinusoid with large wavenumber $q_{x0} = \frac{1}{\sqrt{2}}$ associated to the short-scale linear solution for $\mu=0$, and small amplitude, namely: $Z(t=0)=0.1 \sin ( q_{x0} x)$. Convergence study was performed to confirm that the radial resolution was sufficiently precise not to affect the results.
The - time averaged - radial profile of ZF shear is shown for different values of the normalized ZF friction ${\colr \mu= 10^{-3}, 0.1,0.2}$ [Fig.\ref{fig:fig3}]. \newline
\indent The radial wavenumber is estimated from the time-averaged profiles - normalized to their amplitude - by graphically fitting a square-wave $f(x) = 2\left(2\lfloor q_0 x\rfloor - \lfloor 2 q_0 x \rfloor \right) + 1$, for each value of $\mu$, with a free parameter $q_0$ (wavenumber), where the floor function $\lfloor x \rfloor$ denotes the nearest integer not greater than x. Then, the wavelength is computed as $\lambda = \frac{2\pi}{q_0}$, and the step-size is defined as $\stair= \frac{\lambda}{2 \pi}$.
The associated nonlinear dispersion relation is shown {\colr for 5 values of ZF friction $\mu = 10^{-3},5\times 10^{-3},0.1,0.2$ and $0.25$ }  [Fig.\ref{fig:fig4}].
The following scaling law is obtained:
\begin{equation}
\stair_{\rm stair} \sim \alpha \log\mu + {\rm const}, \quad {\rm with} \quad \alpha \simeq -0.41.
\label{scaling-num}
\end{equation}
This scaling is in good agreement with the prediction $\stair_{\rm stair} \sim -0.34\log \mu + {\rm const}$ obtained from a heuristic nonlinear estimation [Fig.\ref{fig:fig2}]. \newline
\indent \emph{Discussion.$-$} Our results indicating $\stair_{\rm stair} \sim \log\mu^{-0.34}$ analytically and $\stair_{\rm stair} \sim \log\mu^{-0.41}$ numerically can be translated to the collisionality dependence since $\mu \propto \nu_{ii}$ in tokamak plasmas. Interestingly, this trend is opposite to the one due to a residual transport in the absence of turbulent avalanches in the traffic-jam model of the zonal staircase, if one takes the residual transport from collisions \cite{KosugaDiamondGurcan2013}. As more experimental data such as those in Ref.\cite{MJChoi2024} become available, one can learn whether the mechanism discussed in this work plays a more dominant role in those plasmas. In addition, global gyrokinetic simulations accompanied by a systematic parameter scan would be desirable.

In summary, we presented a novel mechanism of scale-selection relevant to the $E \times B$ staircase. Scaling laws - derived both analytically and numerically - are obtained, Eqs.(\ref{scaling-ana}) and (\ref{scaling-num}) which predict the staircase step-size $\stair_{\rm stair}$ v.s. zonal flow friction $\mu$.

\section*{Acknowledgements}
The authors would like to thank P.H. Diamond, Minjun Choi, K. Obuse, S. Takehiro, G. Dif-Pradalier, Y. Kosuga, I. Dodin,
Jae-Min Kwon, Lei Qi, Juhyung Kim, Sumin Yi, and Janghoon Seo for helpful discussions,
and feedbacks from participants at the AAPPS 2023 conference in  Nagoya, Japan. This work was supported by R\&D Program through Korean Institute of Fusion Energy (KFE) funded by the Ministry of Science and ICT of the Republic of Korea (KFE-EN2641-12), and by the National Research Foundation of Korea (NRF) funded by the Ministry of Science and ICT of the Republic of Korea (Grant No. 2023R1A2C1007735).

\end{document}